# Depth calibrations of a 2D-CMOS-based partially coherent light interferometer

C. Pernechele[1], S. Chinellato[1,2], F. Manzan[1], S. Carmignato[3], A. Voltan[4]

[1]INAF - National Institute for Astrophysics, Astronomical Observatory of Cagliari (ITALY)
[2]Dipartimento di Astronomia – Università di Padova (ITALY)
[3]DTG – Dipartimento di Tecnica e Gestione dei Sistemi Industriali, Università di Padova (ITALY)
[4]DIMEG - Dipartimento di Innovazione Meccanica e Gestionale, Università di Padova (ITALY)

*claperne@oa-cagliari.inaf.it*

**Abstract**

This work concerns the development of a 3D measuring system able to realize non-contact surface topography with millimetric depth-range and micrometric resolutions both in the spatial and depth axes. The optical concept is based on the well known low coherence interferometry (LCI) technique. The most widespread set-up of such technique is that of measuring only a point at a time with a 2D scanning system that permits the measure on an area. The novel concept of our instrument is based on the use of a 2D sensor (CMOS), where every single pixel measures a point on the object and this permits a fast analysis on square centimeters areas without the need for any precise (and expensive) scanning system. We present here accurate depth calibration which shows the potentiality of this instrument.

## 1   Introduction

The use of low coherence interferometry (LCI), in this context also referred as optical coherence tomography (OCT), as a tool for topographic measurement at micrometric depth accuracy has been largely demonstrated [1]. The LCI is able to analyse the surface roughness with sub-micrometric depth resolution but, in contrast with the classical interferometry, it also resolves the phase ambiguity, thus making it possible to measure the surface topography up to a millimetric scale range. Moreover this technique permits to render visible, depending on the material's transparency at the working wavelength, different stratifications inside the object under study, allowing for a tomographic view. This technique can be more easily used today in several fields of application due to the availability of relatively low cost and stable partially coherent sources (super-luminescent light emitting diodes - SLEDs). These sources have high spatial and low temporal coherences (partial coherence), necessary to make useful this optical principle. Described applications range from medical application [2], artwork diagnostic [3,4], range-finder for astronomical applications [5], fast and precise alignment of optical instrumentation [6] and others. Actually the most widespread optical set-up is the one using fibre interferometer and only one point per unit time is acquired; the 2D scan speed limits the acquisition rate for high-resolution images to a few hertz. A simple way to improve the frame rate of such an imaging system is to use a parallel detection scheme which allows one to remove the transverse scanner used in standard OCT setups and to acquire a complete image during only one depth scan. Two-dimensional CCD cameras have already been used as detection devices for this purpose [7]. Our instrument uses a 2D CMOS sensor as described in [8].

## 2     Instrument description

As extensively described in [8], our chosen optical set up is a classical Michelson interferometer with a 830 nm (30 μm of coherence length) SLED as source. The image is formed by means of a telecentric objective and is recorded on a CMOS image sensor of 1280 x 1024 pixels of 5 μm square. The spatial resolution is of 13 μm/px. In the actual configuration the elaboration is made in a PC placed outside the CMOS board and the speed of sampling is therefore limited by the USB2 connection. When downloading the full frame the read-out speed is of 15 fps (frame per second) and an image is acquired every 0.1 μm in the depth dimension. The elaboration is performed almost in real time and therefore a sample depth range of 150 μm is measured in 100 seconds.

## 3     In-depth calibration

In order to calibrate the instrument in the depth direction, measures have been taken at the Laboratory of Industrial and Geometrical Metrology of the University of Padova. For the calibration of the vertical measurement axis of the device, depth setting standards with depths into the millimeter range are needed. The tests have been done on a flat surface and on a calibrated spherical surface in order to set the limit slope angle of the measurement.

### 3.1 Tests on flat surface

In order to test the depth precision of the instrument we measured an optical flat surface, with calibrated flatness below 0.1 μm. For this test we used a sub-region of the sensor of 256x256 pixels. At first the optical flat was put at the right distance where the fringes appear and then was tilted until only one fringe appears, this assures that the optical flat is orthogonal to the wavefront. The measured rms along a profile (1x256 pixels) is 150 nm and the residuals are shown in Fig. 1 (left panel). We furtherly tilted the optical flat of 11 fringes, the rms is 200 nm and residuals are shown in the right panel of Fig.1. The difference between the two cases is under study.

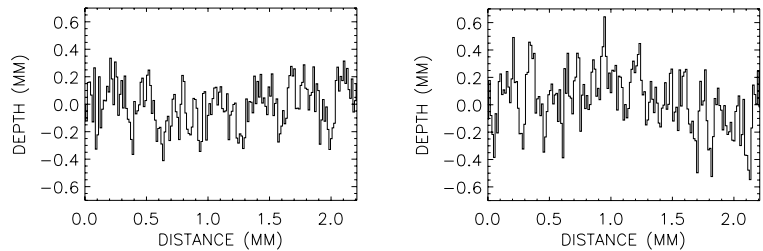

Figure 1: Residuals for a profile: orthogonal (left) and tilted (right) surface with respect to the wavefront.

### 3.2 Limit slope angle

Calibrations have also been made on a calibrated sphere of 5.001 mm of radius in order to set the limit angle and the relative accuracy as shown in Fig. 2.

In this case we used a sub-region of 512x256 pixel in order to visualize the entire sphere's profile. The sphere has a scattering surface and the resulting limit angle is 56°. The precision of the measure with respect to the angle is shown in Fig. 2, where the difference between the measured surface and an ideal one is plotted.

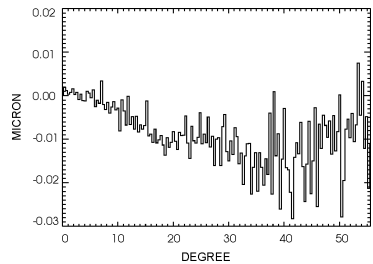

Figure 2: Residual for the limit angle of a sphere.

**4      Conclusions**

We present calibration results for our low coherence interferometer for contactless 3D measurements with a micrometric depth resolution. Calibration on a flat and a curved surface has been performed. Results show that the resolution on the depth axis reaches submicrometric values at least for flat surfaces. The trend of the resolution with respect to the angle has been also mapped. Further calibrations are scheduled.